\documentclass[twocolumn,floatfix,prl,aps,showpacs,superscriptaddress,longbib]{revtex4-2} 
\bibpunct{[}{]}{,}{n}{}{}

\usepackage{graphicx,epsf,amsmath,amssymb}
\usepackage{bm,bbm}
\usepackage{enumerate}

\usepackage[usenames,dvipsnames,svgnames,table]{xcolor}
\usepackage[colorlinks,citecolor=blue,urlcolor=blue]{hyperref}
\usepackage{empheq}
\usepackage{ulem}

\begin{document}

\title{Three-dimensional chiral Veselago lensing}

\author{S. Tchoumakov} 
\email{serguei.tchoumakov@neel.cnrs.fr}
\affiliation{Univ. Grenoble Alpes, CNRS, Grenoble INP, Institut N\'eel, 38000 Grenoble, France}

\author{J. Cayssol}
\affiliation{Univ. Bordeaux, CNRS, LOMA, UMR 5798, F-33405 Talence, France}

\author{A. G. Grushin} 
\affiliation{Univ. Grenoble Alpes, CNRS, Grenoble INP, Institut N\'eel, 38000 Grenoble, France}

\date{\today}

\begin{abstract}
The effect by which light focuses upon entering a medium with a negative refractive index, known as Veselago lensing, may enable optical imaging below the diffraction limit.
Similarly, focusing electrons across a $pn$-junction could realize a technologically promising electronic Veselago lens.
However, its scope remains limited by the lack of three-dimensional platforms and its insensitivity to computational degrees of freedom, like spin or chirality. 
Here we propose a single-material three-dimensional electronic Veselago lens that selectively focuses electrons of a given chirality. 
Using the chiral anomaly of topological semimetals it is possible to create a sharp $pn$-junction for a single chirality, a chiral Veselago lens, and tune it with a magnetic field to an ideal lensing condition.
We estimate that chiral Veselago lensing is observable in non-local transport and spectroscopy experiments. In particular we show that the chiral Veselago lens leads to giant non-local magnetoresistance.
\end{abstract}


\maketitle

{\it Introduction.} 
The similarities between the light-ray construction in optics and the semi-classical trajectories of electrons~\cite{Yan_2020,spector} host the potential for new applications in electronics, such as electronic lenses, interferometers or beam-splitters at the micrometer scale~\cite{Cheianov1252,PhysRevX.7.041026,karalic2019electronhole}.
To reach this potential, controlling the involved material interfaces is crucial. 
In optics, the trajectory of light is deflected at the interface between two media, allowing to focus, guide and disperse light controllably. 
In particular, between media with opposite handedness, where the optical index changes sign, lensing can occur even for a flat interface, a phenomenon coined Veselago lensing~\cite{Veselago_1968}. 
In electronics, a $pn$ junction can act like an electronic Veselago lens by focusing two-dimensional (2D) electrons with pseudo-relativistic, linear band dispersion~\cite{Cheianov1252,PhysRevB.95.214103,Libisch_2017,Tian_2017,Zhang_2017,Zhang_2019,Yang_21}, an effect that has only been realized in graphene~\cite{lee2015observation,Chen1522,zhang2021gatetunable,PhysRevB.100.041401}. 

The prospect of creating useful devices based on electronic Veselago lenses faces three fundamental problems. 
The first problem is that we lack a three-dimensional (3D) Veselago lens. Theoretically, a $pn$ junction sharper than the Fermi wavelength based on 3D relativistic metals, like Weyl or Dirac metals~\cite{Armitage2018}, can Veselago-lens~\cite{PhysRevB.95.214103,Yang_21}, in analogy with graphene. While bulk 3D metals have the advantage of a larger carrier density, this property is also a drawback as they are not easily gated. 
The second problem is creating a clean interface. A $pn$ junction created via chemical doping or by interfacing $p$-doped and $n$-doped samples will likely be disordered by impurities or lattice mismatch. The interface could accumulate charge, affecting electronic transport in undesirable ways~\cite{band_bending_review}.
The last problem is that any proposed or realized electronic Veselago lens seems insensitive to computational degrees of freedom, like spin or chirality.
These three problems materialize in the challenge to realize a single-material 3D Veselago lens that could select specific electronic degrees of freedom.
\begin{figure}[t]
	\centering
	\includegraphics[width=\columnwidth]{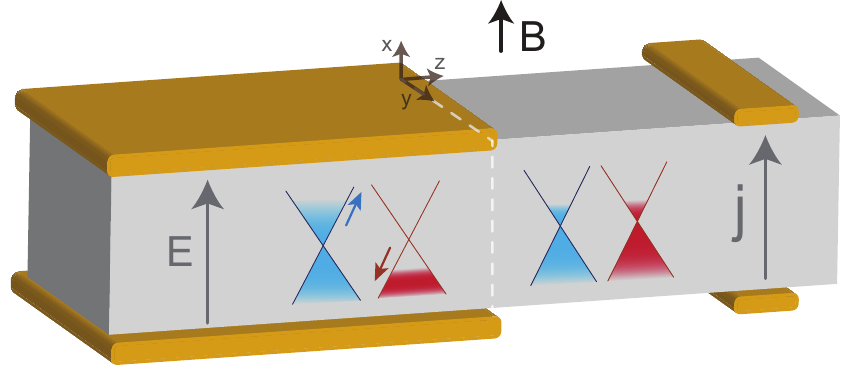}
	\caption{Chiral Veselago lens based on the chiral anomaly. A uniform magnetic field $\mathbf{B}$ is applied to a Weyl semimetal, while the electric field $\mathbf{E}$ is only applied
	for $z < 0$, with two metallic gates (left orange/gold gates). This configuration generates a charge imbalance between Weyl quasiparticles of opposite chiralities (red and blue cones) for $z < 0$ due to the chiral anomaly, while maintaining the total charge constant. This creates an ideal $pn$ junction for a single chirality (blue cones), that Veselago-lenses a measurable non-local current at $z>0$, tuned by changing $\mathbf{B}$, and measured by narrow electrodes (in orange/gold at right).}
	\label{fig:0}
\end{figure}

In this work we take a significant step to solve these problems by proposing how to Veselago-lens a single chirality of 3D Weyl quasiparticles controllably. 
The chirality is a quantum mechanical degree of freedom of 3D Weyl quasiparticles, labelled by $\pm$ depending on whether the spin is aligned or anti-aligned with the electron's momentum, which can be used for computation~\cite{kharzeev2019chiral}.
By locally activating the chiral anomaly, an effect that overpopulates one chirality with respect to the other in the presence of colinear electric and magnetic fields, it is possible to create a sharp $pn$-junction for a single chirality. 
The resulting chiral Veselago lensing can be detected either by spectroscopic probes, like scanning-tunneling microscopy (STM), or non-local transport.

Because of the chiral anomaly, the lensing can be tuned with electric and magnetic fields to modulate the intensity of the image charge in STM or the image current in non-local transport. We also show that ordinary, quadratically dispersing electrons can Veselago-lens stronger than Weyl electrons, albeit without the degree of control offered by the chiral anomaly and suffering from a charging layer at the $pn$ interface. Therefore the chiral anomaly, unique to Weyl semimetals, is the optimal tool to realize a clean 3D chiral Veselago lens.
\begin{figure*}
	\centering
	\includegraphics[width=\textwidth]{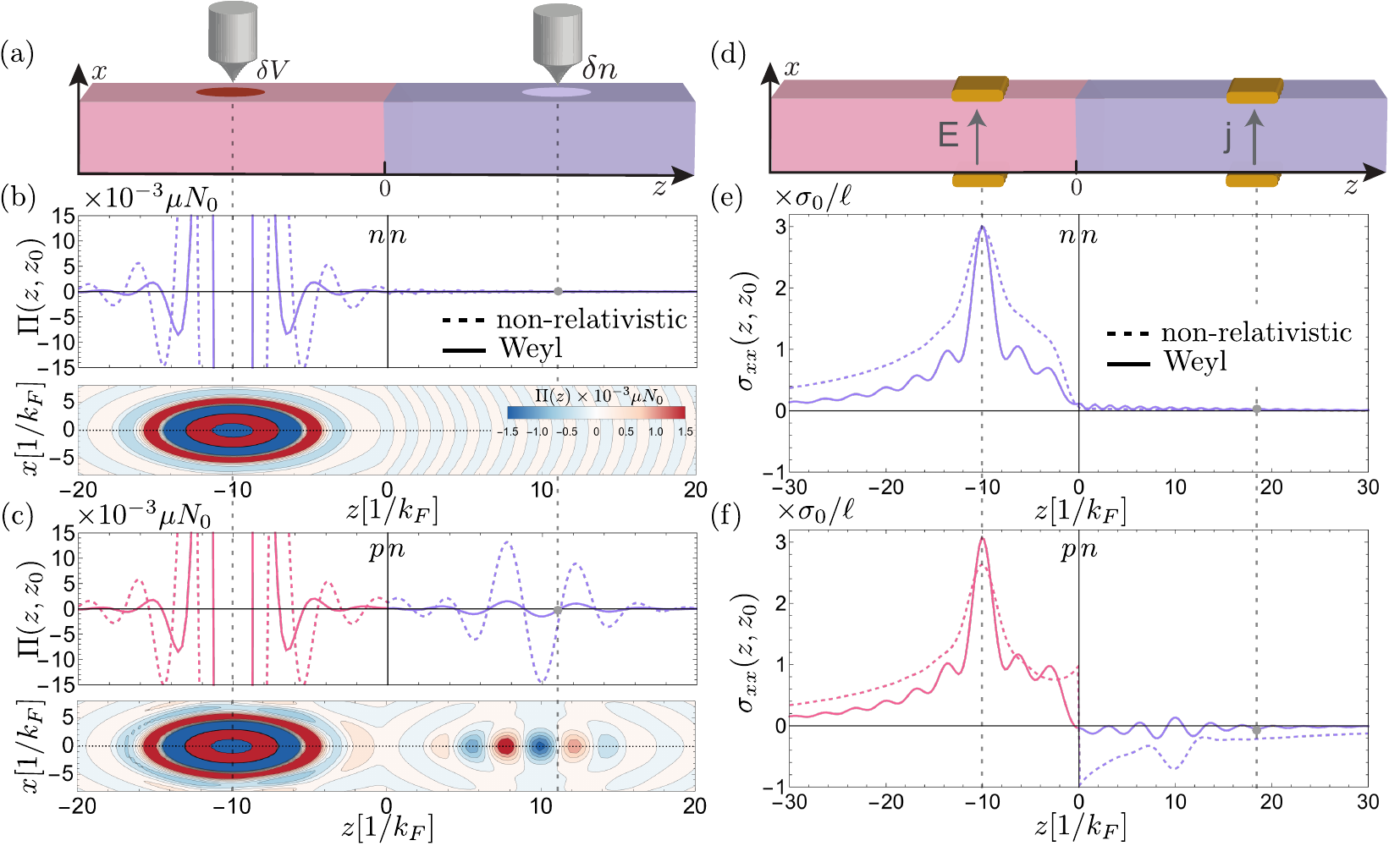}
	\caption{Veselago lensing of a single chirality in spectroscopy and non-local transport.
	(a) Schematic set-up to measure Veselago lensing spectroscopically (see \cite{SuppMat} for a two-chirality set-up). 
	(b) The upper plot shows the non-local polarisability $\Pi(z,z^{\prime})$ for an $nn$ junction as a function of $z$ for $(x,y) = (0,0)$ for both non-relativistic (dashed) and Weyl band structures (solid). The contour plot shows $\Pi(z,z^{\prime})$ in the $(x,z)$ plane for $y = 0$ for the Weyl case.  
	(c) Same as (b) for a $pn$ junction. 
	In (b-c) $\Pi(z,z^{\prime})$ is given in units of $\mu N_0$, with $N_0 = k_F^3/(2\pi^2 \mu_L)$ the density of states on the left of the junction.
	(d) Schematic set-up to measure Veselago lensing in non-local transport (see \cite{SuppMat} for a two-chirality set-up). 
	(e) and (f) show the non-local conductivity $\sigma_{xx}(z,z^{\prime})$ for $nn$ and $pn$ junctions, respectively. 
	(b,e) are calculated with $(m,\mu,\ell)_L = (1,0.5,50)$ and $(v,\mu,\ell)_R = (2,4,50)$ for the non-relativistic case, and $(v,\mu,\ell)_L = (1,1,50)$ and $(v,\mu,\ell)_R = (1,2,50)$ for the Weyl case. (c,f) are calculated with $(m,\mu,\ell)_L = (1,0.5,50)$ and $(v,\mu,\ell)_R = (-1,-0.5,50)$ for the non-relativistic case, and $(v,\mu,\ell)_L = (1,1,50)$ and $(v,\mu,\ell)_R = (1,-1,50)$ for the Weyl case.  
	In (e-f) the conductivity is given in units of the total conductivity $\sigma_0$ over the mean-free path $\ell$ (see~\cite{SuppMat}).}
	\label{fig:1}
\end{figure*}

{\it Model.} Weyl semimetals host pairs of linearly dispersing bands separated in momentum space, known as Weyl cones. Each cone can be described by the Hamiltonian
\begin{align}
	\label{eq:hd}
	\hat{H}_W =(v_F{\bf k} - {\bf b}) \cdot \hat{\boldsymbol{\sigma}} - \mu,
\end{align}
where $\boldsymbol{\sigma}$ is a vector of Pauli matrices, $\mu$ is the chemical potential, $v_F$ is the Fermi velocity, and ${\bf b}$ determines the momentum space separation between Weyl cones. The eigenstates of $\hat{H}_W$ depend on the sign of $v_F$, which defines the chirality $\mathrm{sgn}(v_F)=\chi = \pm$. Weyl cones come in pairs of opposite $\chi$ and, in certain materials, they can be tilted~\cite{SGD15} or anisotropic~\cite{Armitage2018}. Here we focus on materials where these effects are negligible, as discussed at the end.

In presence of electric ($\mathbf{E}$) and magnetic (${\bf B}$) fields, the transport of Weyl quasiparticles depends on their respective chirality, as reflected by the continuity equations
\begin{align}
    \frac{\partial n_{\chi}}{\partial t} + \boldsymbol{\nabla}\cdot{\bf j}_{\chi} = \chi\frac{ e^2}{2\pi^2 \hbar^2}\mathbf{E}\cdot {\bf B},
    \label{eq:chiranom}
\end{align}
where $n_{\pm}$ and ${\bf j}_{\pm}$ are, respectively, the charge and current distributions of carriers with $\chi = \pm$. The source term, on the right-hand side, creates an imbalance $\Delta n$ between the chiral charges $n_+$ and $n_-$, without altering the total charge $n = n_+ + n_-$. This term, known as the chiral anomaly, is responsible for anomalous transport responses, such as the enhancement of the magneto-conductance of Weyl semimetals~\cite{Landsteiner2016,Armitage2018}.

The chiral anomaly competes with inter-valley scattering, leading to a charge imbalance between Weyl nodes with opposite chiralities in the steady-state. The inter-valley scattering time $\tau$ can range from a picosecond in TaAs to a nanosecond in Na$_3$Bi~\cite{PhysRevX.4.031035,Behrends2016}.  The chiral charge imbalance $\Delta n = n_+ - n_- = \frac{\tau e^2}{2\pi^2 \hbar^2}\mathbf{E}\cdot {\bf B}$ equals the carrier density $n$ for a magnetic field $B_c \equiv 2\mu^3/(3v^3 e^2 \tau E)$. For $n = 10^{21}$~cm$^{-3}$, $\tau = 10^{-9}$~s~\cite{Behrends2016,PhysRevB.89.245121,PhysRevX.4.031035} and $E \approx 10^{6}$~V.m$^{-1}$, this crossover field is $B_c = 1$~T. Beyond $B_c$, the chiral anomaly reverses the type of carriers of one of the Weyl nodes, e.g. from electrons to holes. Hence, if $B>B_c$ is applied homogeneously and the $E$ is not homogeneous, e.g. stepwise as illustrated in Fig.~\ref{fig:0}, the chiral anomaly creates an effective $pn$ junction for a single Weyl chirality.

The main goal of this work is to explain how to realize and detect a single-chirality Veselago lens, under the above conditions. To this end, we discuss spectroscopy and non-local transport responses of a single Weyl cone to determine the ideal conditions for single-chirality Veselago lensing. Combining the contributions from the two chiralities we discuss how chiral Veselago lensing can be detected using the simple set-up in Fig.~\ref{fig:0}. In the Supplemental Material~\cite{SuppMat} we propose two related, albeit less practical, devices to detect directly chiral Veselago lensing in spectroscopy and in non-local transport. 

We model the change in the carrier density with a step-wise chemical potential, where $\mu_L = \left( \mu^3 \pm \frac32 v^3 e^2 \tau \mathbf{E}\cdot{\bf B}\right)^{1/3}$ for $z < 0$~\cite{PhysRevB.93.075114} and $\mu_R = \mu$ for $z > 0$ ($L$ and $R$ denote left and right of the interface, respectively).

\textit{Spectroscopic signatures of chiral Veselago lensing.}
To visualize and understand the ideal conditions for Veselago lensing, we calculate the charge distribution created by a local potential due to an impurity, or the tip of a STM. 
For a local potential $V({\bf r}) = \delta({\bf r}_{\parallel})\delta(z - z_0) \delta V$ there is a redistribution of charge $\delta n (z) = \Pi(z,z_0) \delta V$ that depends on the polarisability ~\cite{giuliani_vignale_2005}
\begin{align}
	\Pi(z,z^{\prime}) = - \frac{1}{2\pi}\int d\omega~ {\rm Tr}\left( \hat{G}(z,z^{\prime}) \hat{G}(z^{\prime},z) \right),
\end{align}
written in terms of the Green's function $\hat{G}(z,z^{\prime})$, where we account for scattering with a mean free path $\ell \gg 1/k_F$ (see \cite{SuppMat} for details). 

In Fig.~\ref{fig:1}(b,c) we plot the polarisability as a function of the $x$ and $z$ coordinates. 
The coordinates are given in units of $1/k_F$, and the polarisability in units of $\mu_L N_0$, with $N_0 = k_F^3/(2\pi^2\mu_L)$ being the density of states at $z<0$, where the impurity is located.
We compare the case of a single Weyl chirality with a 3D electron gas, with a quadratic band structure~\cite{SuppMat}.
In a $nn$ (or $pp$) junction, the charge distribution shows the usual Friedel oscillations on both sides of the junction, see Fig.~\ref{fig:1}(b). 
The period is different on both sides due to the change in the Fermi wavevector when crossing the interface~\cite{giuliani_vignale_2005}. 

The main difference between the $nn$ junction in Fig.~\ref{fig:1}(b) and the $pn$ junction in Fig.~\ref{fig:1}(c) is the appearance of an image charge as a consequence of Veselago lensing. Veselago lensing occurs because the in-plane group velocity changes sign across the interface, ${\bf v}_L = - {\bf v}_R$, due to the conservation of the in-plane wavevector ${\bf k}_{\parallel}$. This condition can be met for both a 3D electron gas and a Weyl semimetal $pn$ junction, but the intensity of the image charge is larger for the former than the latter due to the slower decay of Friedel oscillations in a normal electron gas~\cite{PhysRevB.92.241103,PhysRevB.92.224435,PhysRevB.99.165302,PhysRevB.102.205202}. 

However, a drawback of using a 3D electron gas is that the electron density accross the $pn$ junction is not constant, and one should expect a built-in interface potential over a finite distance $d$. When $k_F d \gg 1$, we find~\cite{SuppMat} that the amplitude of the Veselago image rapidly decays~\cite{Chen1522}.
For typical 3D metals $d > 10~\mu$m so $ k_F d \gg 1$~\cite{pierret1996semiconductor} implying that the corresponding image charge is unobservable in practice~\cite{Chen1522}. 
In contrast, in the chiral Veselago lens in Fig.~\ref{fig:0} the total charge remains constant, charges are only transferred between cones of opposite chirality. The electric field extends beyond the parallel plates on a distance of the order of the distance between the two plates, so we can expect a sharp interface potential for thin film samples.

So far we have assumed symmetric $pn$ junctions, i.e. those where $k_F$ is equal on both sides of the interface. 
Deviations from this condition blur and change the location of the image charge~\cite{SuppMat}. 
Reaching ideal lensing is unrealistic with 3D electron gases because finding junctions with equal $k_F$ but opposite carrier types is challenging in practice. Later we will argue how the chiral anomaly of 3D Weyl semimetal aids to tune into this ideal condition by varying the magnetic field.

\textit{Non-local transport signatures of chiral Veselago lensing.} 
While the image charge is convenient to understand how Veselago lensing can be enhanced (see Supplemental Material~\cite{SuppMat} for a proposed device), surface states contributions must be factored out~\cite{Inoue1184,Batabyale1600709,Kourtis16,Yuaneaaw9485} to reveal Veselago lensing. 
A bulk, non-local transport measurement, such as that depicted in Fig.~\ref{fig:1}(d), is in this sense a simpler set-up. In a local electric field ${\bf E}(z) = {\bf E}_0\delta(z - z_0)$ the electronic current $j_{\mu}(z) = \sigma_{\mu\nu}(z,z_0) E_{\nu}$ is obtained from the non-local conductivity~\cite{PhysRevB.51.10085,BarangerStone89}
\begin{align}
    \sigma_{\mu\nu}(z,z^{\prime}) = \int \dfrac{dS_z dS_{z^{\prime}}}{\pi \mathcal{A}}{\rm Tr}\left( \hat{j}_{\mu} {\rm Im}\hat{G}(z,z^{\prime})\hat{j}_{\nu} {\rm Im}\hat{G}(z,z^{\prime})\right),
    \label{eq:sigma}
\end{align}
where $\hat{j}_{\mu}$ are the components of the current operator, with $\mu = x,y,z$, and $S_z$ and $S_{z^{\prime}}$ are planes at $z$ and $z^{\prime}$ with areas $\mathcal{A}$. The non-local conductivity is a complex quantity, that accounts for the dephasing between the two probes. In Eq.~\eqref{eq:sigma} we only show its real part, which can be measured by averaging the conductivities obtained after permuting the positions $z$ and $z^{\prime}$ of the leads (see Supplemental Material~\cite{SuppMat}).

In Figs.~\ref{fig:1}(e,f), we show the non-local conductivity for a $nn$ (or $pp$) and a $np$ junction. For a $nn$ junction, the current is positive and decays exponentially away from the input electric field, as shown in Fig.~\ref{fig:1}(e). In contrast, for the $pn$ junction, the current changes sign when crossing the interface and its magnitude peaks close to the image charge, signaling the presence of a Veselago lens~\cite{lagasse}. 
Note that the negative current does not violate energy conservation, since the total current is positive, and is a consequence of the change of type of the main carriers. 

As with the image Veselago charge, the image Veselago current is larger for a 3D electron gas than for a Weyl semimetal, where it oscillates close to zero, but negative on average. These oscillations are absent in the 3D electron gas which is a single-band model, and thus we attribute them to interband excitations that lead to hole-like and electron-like regions. Lastly, we find that ideal lensing is achieved when both sides of the $pn$ junction are tuned to the have the same $k_F$, as for the image Veselago charge.
\begin{figure}[t]
	\centering
	\includegraphics[width=\columnwidth]{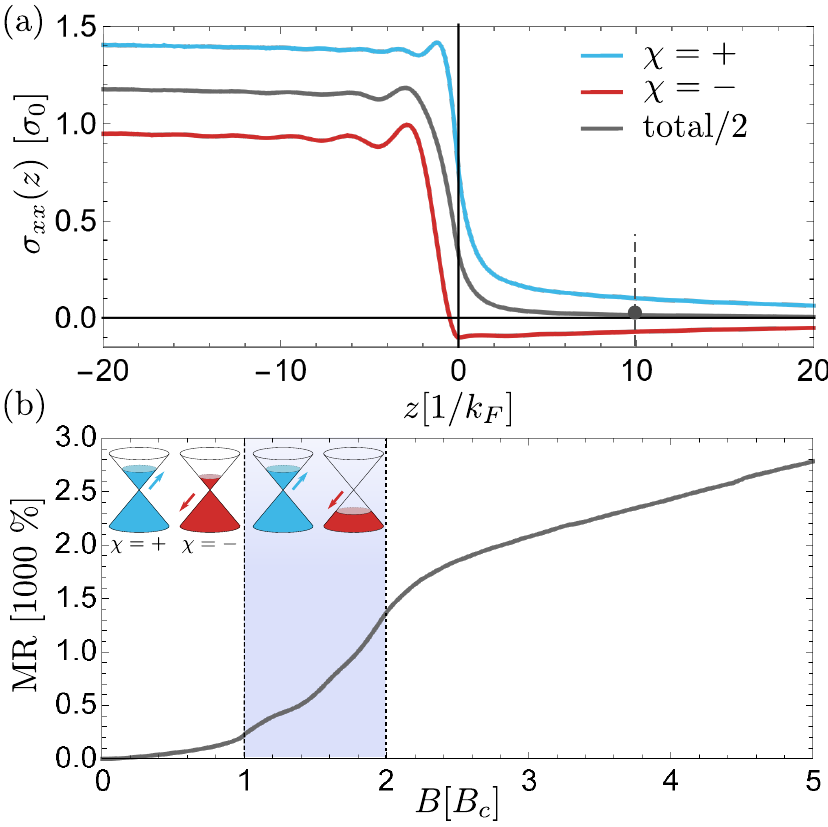}
	\caption{Transport signatures of the chiral Veselago lens depicted in Fig.~\ref{fig:0}. 
	(a) The total non-local conductivity $\sigma_{xx}(z)$ (dark curve) for $B = 2B_c$. The contribution of each chirality is plotted as colored curves.  
	$\sigma_{xx}(z)$ is uniform for $z < 0$ where the electric field is constant and drops for $z > 0$, with a negative conductivity for Weyl quasiparticles with chirality $\chi = -$. 
	(b) Giant magnetoresistance due to the chiral anomaly for a probe located at $z = 10/k_F$; see dashed line in (a). The magnetic field is in units of $B_c = 2\mu^3/(3v^3 e^2 \tau E)\approx 1T$ (see text). Chiral Veselago lensing starts at $B=B_c$, which marks the change from electrons to holes for chirality $\chi=-$ (inset red cone). For $B>B_c$ carriers with $\chi=-$ contribute with a negative current that is maximal at $B = 2B_c$, when the Fermi wavevector is opposite of both sides of the junction. The strong increase in magnetoresistance,  marked by the blue region, is thus a fingerprint of chiral Veselago lensing.}
	\label{fig:2}
\end{figure}

{\it 3D Chiral Veselago lens.} 
We have previously considered the transport of each chirality individually, while in the set-up of Fig.~\ref{fig:0}, the two chiralities are present. 
In this set-up the electric field $E$ is finite for $z < 0$, and zero for $z > 0$.
In Fig.~\ref{fig:2}(a) we show the resulting total non-local conductivity and the contribution of each chirality, calculated using the expression
\begin{align}
    \sigma_{xx}(z) = \int_{-\infty}^{0} dz^{\prime}~\sigma_{xx}(z,z^{\prime}).
\end{align}
For $z > 0$, $\sigma_{xx}(z)$ decays slowly and the contribution of each chirality is about $10\%$ of the bulk conductivity at $z k_F = 10$. Since the conductivity is negative for one chirality and positive for the other, the two contributions compensate and lead to a giant magnetoresistance with respect to the configuration without chiral anomaly. This follows from our previous discussion: because of the chiral anomaly, carriers of one chirality (here $\chi = -$) experience a $pn$ junction while carriers of the opposite chirality (here $\chi = +$) experience a $nn$ junction. Since $E$ extends for all $z < 0$, the $pn$ junction leads to an extended, rather than localized, Veselago image. Nonetheless, the ideal condition for an optimal image is still $k_{F,L} = k_{F,R}$. Accordingly, we find that the negative current is maximal when this condition is met.

The total $\sigma_{xx}(z)$ of the junction is a sum of the contributions of each Weyl cone. As seen in Fig.~\ref{fig:2}(a), the total $\sigma_{xx}(z)$ is positive throughout the junction (dark curve), since the total charge is conserved, hiding the signature of Veselago lensing. However, Veselago lensing becomes apparent when studying the non-local resistivity as a function of $B$, as shown in Fig.~\ref{fig:2}(b). At $B<B_c$ there is no chiral Veselago lensing and the magnetoresistance increases moderately. This is due to the increasing mismatch between the carrier densities on each side of the junction. Once $B > B_c$, an effective $pn$ junction is formed for one chirality (here $\chi = -$) that leads to a negative contribution to current and thus to a giant increase in the magnetoresistivity (shaded region in Fig.~\ref{fig:2}(b)). The amplitude of the negative current is maximal at $B = 2B_c$ where  $k_{F,L} = k_{F,R}$ for $\chi = -$, and Veselago lensing is optimal. For $B > 2B_c$ the amplitude of the negative current decreases, and the magnetoresistance increases moderately. The strong increase in magnetoresistance within $[B_c,2B_c]$, highlighted in Fig.~\ref{fig:2}(b), is thus a fingerprint of chiral Veselago lensing.

{\it Discussion.} We have determined how the chiral anomaly of Weyl semimetals can be used to create a Veselago lens that distinguishes electrons by their chirality. We have uncovered that the chiral Veselago lens based on Weyl semimetals is advantageous for several reasons. First, a chiral Veselago lens eliminates the charge build up at the interface, since it requires a single material, and the total charge remains constant along the sample. Second, the chiral anomaly allows to control the $k_F$ of each chirality using a magnetic field. This allows to detect chiral Veselago lensing by measuring non-local conductivity and tune it to the ideal condition for lensing. This property goes beyond current graphene-based proposals, since the chiral anomaly is unique to 3D Weyl semimetals.

We have neglected several effects in our computations. First, we neglected the orbital localization in a magnetic field, since this occurs over the magnetic length scale $\ell_B \approx 25.6$ nm$/\sqrt{B {\rm [T]}} \gg z,\lambda_F$. Second, tilted Weyl cones~\cite{SGD15} could prevent a perfect matching of $k_F$. However, the chiral anomaly is often observed in systems where the Weyl cones are generated by magnetic field from a lightly doped semimetal, e.g. GdPtBi, ZrTe$_5$, or KZnBi~\cite{GdPtBiPhysRevB.95.161306,KZnBiPhysRevX.11.021065,Armitage2018,PhysRevB.85.195320} where the effect of tilting is negligible. Lastly, Fermi arcs will be negligible in non-local transport, as it is a bulk probe.

The estimates we provide indicate that chiral Veselago lensing is observable under moderate electric and magnetic fields. Non-local transport experiments have probed the chiral anomaly, albeit without $pn$-junctions and lensing~\cite{PhysRevX.4.031035,zhang2017room}, suggesting that our proposal can be realized using current technology. Our work opens the possibility to realize a device that can control a computational degree of freedom using Veselago lensing in 3D materials, enabled by the chiral anomaly.

\bigskip

{\it Acknowledgements.-} We thank B. Gotsman, H. Schmid, A. Molinari and J. Gooth for discussions about experimental details.
A. G. G and S. T acknowledge financial support from the European Union Horizon 2020 research and innovation program under grant agreement No 829044 (SCHINES).
A. G. G. is also supported by the ANR under the grant ANR-18-CE30-0001-01 (TOPODRIVE).
 
\bibliographystyle{apsrev4-1}
\bibliography{bibliography}

\clearpage
\appendix
\begin{widetext}

\renewcommand{\thefigure}{S\arabic{figure}}
\setcounter{figure}{0} 
\setcounter{secnumdepth}{2} 

\section{Green's function of a junction between two non-relativistic electron gas}

We consider a normal electron gas without spin where the dynamics are described by the Schrödinger Hamiltonian
\begin{align}
    \hat{H} = \frac{k^2}{2m} - \mu,
    \label{app:eq:hs}
\end{align}
which depends on two parameters: the band mass $m$ and the chemical potential $\mu$. Note that in this equation we have set units $\hbar = 1$.

Since the mass term is also position dependent, we symmetrize this Hamiltonian and use \cite{mass1,mass2}
\begin{align}
    \hat{H} = -\frac12\nabla\left(\frac{1}{m(z)} \nabla\right) - \mu.
\end{align}
The current operator is defined as
\begin{align}
    \hat{\bf J} = \frac{e\hbar}{2mi}\psi^{\dagger}({\bf r})\left( \nabla - \nabla^{\dagger} \right)\psi({\bf r}).
\end{align}

\subsection{Bulk Green's function}

It is useful to write three forms of the Green's function: in momentum space $G^{\infty}(i\omega_n,{\bf k})$, in real space $G^{\infty}(i\omega_n,{\bf x},z)$ and in mixed representation $G^{\infty}(i\omega_n, {\bf k}_{\parallel}, k_z)$. We define the mixed representation by considering momentum space along the interface and real space transverse to it. The $\infty$ symbol on the Green's functions is used to remind that it correspond to the solution for a bulk sample, where parameters are constant over all space. 
\begin{enumerate}[(i)]
    \item in momentum space
    \begin{align}
        G^{\infty}(i\omega_n, {\bf k}; m,\mu) = \frac{1}{\frac{q^2}{2m} - \mu - i\omega_n},
    \end{align}
    \item in real space
    \begin{align}
        G^{\infty}(i\omega_n, {\bf r}; m,\mu) &= \frac{1}{(2\pi)^3} \int_{0}^{\infty} dq q^2 ~\int_{0}^{\pi} d\theta \sin(\theta)~d\phi \frac{e^{i q r \cos(\theta)}}{\frac{q^2}{2m} - \mu - i\omega_n} = \frac{m}{2\pi^2} \int_{0}^{\infty} dq q^2 ~\int_{-1}^{1} du~ \frac{e^{i q r u}}{q^2 - q_0^2},\\
        &= \frac{-im}{2\pi^2 r} \int_{0}^{\infty} dq q ~ \frac{e^{i q r}}{q^2 - q_0^2} = \frac{m}{2\pi r} e^{i  {\rm sgn}(m\omega_n) q_0 r},
    \end{align}
    where $q_0 = \sqrt{2m(\mu + i\omega_n)}$.
    \item in mixed representation 
    \begin{align}
        G^{\infty}(i\omega_n, {\bf k}_{\parallel}, z; m,\mu) = \frac{1}{2\pi} \int_{0}^{\infty} dq \frac{e^{i q r}}{\frac{q^2}{2m} + \frac{k_{\parallel}^2}{2m} - \mu - i\omega_n} = \frac{m}{\pi} \int_{0}^{\infty} dq \frac{e^{iqr}}{q^2 - q^2(k_{\parallel})} = \frac{i{\rm sign}(m \omega_n) m}{q(k_{\parallel})} e^{i {\rm sgn}(m \omega_n) q(k_{\parallel}) r},
    \end{align}
    where $q({\bf k}_{\parallel}) = \sqrt{2m(\mu + i\omega_n) - k_{\parallel}^2}$.
\end{enumerate}

\subsection{Junction Green's function}

In the situation of a junction where parameters change abruptly from $(m_L,\mu_L)$ at $z < 0$ to $(m_R,\mu_R)$ at $z > 0$, one can decompose the corresponding Green's function over the bulk Green's function to satisfy the boundary conditions. The boundary conditions are obtained by integrating the Schrödinger equation on a small segment around the interface and necessitate the continuity of the Green's function to be well defined
\begin{align}
    &G_{L}(i\omega_n, {\bf k}_{\parallel}, z, 0) = G_{R}(i\omega_n, {\bf k}_{\parallel}, z, 0),\\
    &\frac{1}{m_1} \partial_{z^{\prime}} G_{L}(i\omega_n, {\bf k}_{\parallel}, z, 0) = \frac{1}{m_2} \partial_{z^{\prime}} G_{R}(i\omega_n, {\bf k}_{\parallel}, z, 0).
    \label{app:eq:bc1}
\end{align}
where we introduce the Green's functions $G_{L}(i\omega_n, {\bf k}_{\parallel}, z, z^{\prime})$ and $G_{R}(i\omega_n, {\bf k}_{\parallel}, z, z^{\prime})$ defined for $z^{\prime} < 0$ and $z^{\prime} > 0$ respectively, \textit{i.e.} left and right side of the junction. In order to compute the Green's functions that satisfy the boundary condition~\eqref{app:eq:bc1}, we expand the Green's functions $G_{L}$ and $G_{R}$ over the Green's functions $G_{L}^{\infty}$ and $G_{R}^{\infty}$ of the corresponding bulk material, \textit{i.e.} the Green's functions where parameters in Eq.~\eqref{app:eq:hs} are constants equal to ($m_L,\mu_L$) and ($m_R, \mu_R$) respectively. For example, for $z < 0$ we write
\begin{align}
       &G_{L}(i\omega_n, {\bf k}_{\parallel}, z, z^{\prime}) = G^{\infty}_{L}(i\omega_n, {\bf k}_{\parallel}, z - z^{\prime}) + R G^{\infty}_L(i\omega_n, {\bf k}_{\parallel}, z + z^{\prime}),\\
       &G_{R}(i\omega_n, {\bf k}_{\parallel}, z, z^{\prime}) = T G^{\infty}_R(i\omega_n, {\bf k}_{\parallel}, z - z^{\prime}),
\end{align}
with $R$ and $T$ constants in $z^{\prime}$ but functions in $z$. The boundary condition~\eqref{app:eq:bc1} then implies
\begin{align}
   &G^{\infty}_{L}(i\omega_n, {\bf k}_{\parallel}, z) + R G^{\infty}_L(i\omega_n, {\bf k}_{\parallel}, z) = T G^{\infty}_R(i\omega_n, {\bf k}_{\parallel}, z)\\
   &\frac{1}{m_L}\partial_{z}G^{\infty}_L(i\omega_n, {\bf k}_{\parallel}, z) - R \frac{1}{m_L}\partial_{z}G^{\infty}_L(i\omega_n, {\bf k}_{\parallel}, z) = T \frac{1}{m_R}\partial_{z} G^{\infty}_R(i\omega_n, {\bf k}_{\parallel}, z),
\end{align}
which leads to
\begin{align}
    R &= \frac{J_L^{\infty}/G_L^{\infty} - J_R^{\infty}/G_R^{\infty}}{J_L^{\infty}/G_L^{\infty} + J_R^{\infty}/G_R^{\infty}}, \\
    T &= \frac{2J_L^{\infty}/G_R^{\infty}}{J_R^{\infty}/G_R^{\infty} + J_L^{\infty}/G_L^{\infty}},
\end{align}
where we introduce the current function $J^{\infty} = \frac{1}{m} \partial_z G^{\infty}(i\omega_n, {\bf k}_{\parallel}, z)$. More explicitly this leads to reflection and transmission coefficients that remind of Fresnel coefficients in optics
\begin{align}
    &R = \frac{q_L(k_{\parallel})/|m_L| - q_R(k_{\parallel})/|m_R|}{q_L(k_{\parallel})/|m_L| + q_R(k_{\parallel})/|m_R|},\\
    &T = \frac{2 q_R(k_{\parallel})/|m_R|}{q_L(k_{\parallel})/|m_L| + q_R(k_{\parallel})/|m_R|} e^{i {\rm sign}(\omega_n)({\rm sign}(m_L)q_L(k_{\parallel}) - {\rm sign}(m_R)q_R(k_{\parallel})) |z|},
\end{align}
with $q_{i}(k_{\parallel}) = \sqrt{2m_i(\mu_i + i\omega_n) - k_{\parallel}^2}$. 

We obtain the Green's function for $z > 0$ following the same procedure and obtain similar expression with a permutation of the $L \leftrightarrow R$ indices.

We compute Friedel oscillations and non-local conductivity by performing a numerical Fourier transform of the above Green's function, which is in momentum space, to have the Green's function in real space. In order to simplify the calculation, we perform part of the Fourier transform analytically, using rotation invariance, for example for $z < 0$ we have
\begin{align}
   &G_{L}(i\omega_n, {\bf r}, z, z^{\prime}) = G^{\infty,1}(i\omega_n, {\bf r}, z - z^{\prime}) + \frac{1}{2\pi} \int dk_{\parallel}k_{\parallel} R G^{\infty,1}(i\omega_n, {\bf k}_{\parallel}, z + z^{\prime}) J_{0}(k r_{\parallel}),\\
   &G_{R}(i\omega_n, {\bf r}, z, z^{\prime}) =  \frac{1}{2\pi} \int dk_{\parallel}k_{\parallel} T G^{\infty,2}(i\omega_n, {\bf k}_{\parallel}, z - z^{\prime}) J_{0}(k r_{\parallel}).
\end{align}

\subsection{Surface states}

We can look for the surface states solution as poles of the Green's function and, more specifically, as poles of the reflection and transmission coefficients. The denominators of $R$ and $T$ are proportional to
\begin{align}
    D(i\omega_n,k_{\parallel}) = \sqrt{2m_L(\mu_L + i\omega_n) - k_{\parallel}^2}/|m_L| + \sqrt{2m_R(\mu_R + i\omega_n) - k_{\parallel}^2}/|m_R|,
\end{align}
and the poles are obtained as a solution to $D(\omega + i 0^+, k_{\parallel}) = 0$. The solution is obtained only if the argument of both square roots is negative and with opposite imaginary parts, that is if we satisfy the two conditions
\begin{enumerate}[$\bullet$]
    \item ${\rm sgn}(m_L m_R) < 0$: we have a $pn$ junction,
    \item $2m_L(\mu_L + \omega)< k_{\parallel}^2$ and $2m_R(\mu_R + \omega) < k_{\parallel}^2$: which does not imply any condition on $\mu_L$ and $\mu_R$.
\end{enumerate}
Then one obtains
\begin{align}
    &\frac{k_{\parallel}^2 - 2m_L(\mu_L+\omega)}{m_L^2} = \frac{k_{\parallel}^2 - 2m_R(\mu_R+\omega)}{m_R^2}\\
    \implies &\omega_{k_{\parallel}} = \frac12 \bigg(\frac{1}{m_L} + \frac{1}{m_R}\bigg) k_{\parallel}^2 - \frac{\mu_R/m_R - \mu_L/m_L}{1/m_R - 1/m_L}.
\end{align}
We observe that we have a flat band behaviour (no dispersion of the surface states, diverging density of states) if $m_L = - m_R$, which is the case of perfect Veselago lensing, so the surface states do not disperse. Note that the effective chemical potential of surface states is an average of the chemical potential on each side of the interface. 

The consistency condition for a pole in $R$ and $T$ is valid only for in-plane wavevector such that
\begin{align}
     k_{\parallel}^2 > {\rm min}\bigg( m_L \frac{\mu_L/m_L - \mu_R/m_R}{1/m_L - 1/m_R} - 2\mu_L , -m_R \frac{\mu_L/m_L - \mu_R/m_R}{1/m_L - 1/m_R}  - 2\mu_R \bigg),
\end{align}
so the spectrum of surface states has a hole at small wavevectors. Also the decay of surface states in each bulk is given by $q_{1,2}(\omega) = \sqrt{2m_{1,2}(\mu_{1,2} + \omega_{k_{\parallel}} + i0^{+}) - k_{\parallel}^2}$, with opposite signs in each bulk.

\section{Green's function for a junction between two Weyl semimetals}

We consider a Dirac electron gas with Hamiltonian
\begin{align}
    \hat{H}_W  = \sum_{i = 1}^{3} v_i (k_i - \delta k_i) \hat{\sigma}_i - \mu,
    \label{app:eq:hd}
\end{align}
which depends on the three Dirac velocities $v_x,v_y,v_z$, the three momentum shifts $\delta k_x, \delta k_y, \delta k_z$ and the chemical potential $\mu$. Note that in this equation we relate units of time and energy by setting $\hbar = 1$. Here, we define the current operator as
\begin{align}
    \hat{J} = e v_F \Psi^{\dagger}({\bf r}) \hat{\sigma}\Psi({\bf r}).
\end{align}

\subsection{Bulk Green's function}

It is useful to write three forms of the Green's function: in momentum space $G^{\infty}(i\omega_n,{\bf k})$, in real space $G^{\infty}(i\omega_n,{\bf x},z)$ and in mixed representation $G^{\infty}(i\omega_n, {\bf k}_{\parallel}, k_z)$. As for the 3D electron gas  we define the mixed representation by considering momentum space along the interface and real space transverse to it. The $\infty$ symbol on the Green's functions is used to remind that it correspond to the solution for a bulk sample, where parameters are constant over all space. 
\begin{enumerate}[(i)]
    \item in momentum space
    \begin{align}
        G^{\infty}(i\omega_n, {\bf k}; m,\mu) = \frac{\mu + i\omega_n + \hat{H}_{0}}{\sum_{i} h_i^2 - (i\omega_n + \mu)^2}.
    \end{align}
    \item in real space 
    \begin{align}
        G^{\infty}(i\omega_n, {\bf r}) &= \frac{1}{(2\pi)^3} \int d^3{\bf k}~\frac{\mu + i\omega_n + \hat{H}_{0}}{\sum_{i} h_i^2 - (i\omega_n + \mu)^2} e^{i{\bf k}\cdot{\bf r}}\\
        &= (\mu + i\omega_n) I_0({\bf r}) \hat{\mathbbm{1}} + v_x( -i\partial_x - \delta k_x)I_{0}({\bf r}) \hat{\sigma}_x + v_y( -i\partial_y - \delta k_y)I_{0}({\bf r}) \hat{\sigma}_y - i v_z \partial_z I_0({\bf r}) \hat{\sigma}_z,
    \end{align}
    with
    \begin{align}
        I_0({\bf r}) &= \frac{1}{(2\pi)^3} \int d^3{\bf k}~\frac{ e^{i{\bf k}\cdot{\bf r}}}{\sum_{i} h_i^2 - (\mu + i\omega_n)^2} = \frac{e^{i(\delta k_x x + \delta k_y y)}}{4\pi|v_x v_y v_z|\tilde{r}} e^{i{\rm sgn}(\omega_n)(\mu + i\omega_n)\tilde{r}},
    \end{align}
    where $\tilde{r} = \sqrt{(x/v_x)^2 + (y/v_y)^2 + (z/v_z)^2}$. From this we obtain,
    \begin{align}
        G^{\infty}(i\omega_n, {\bf r}) = \left[ (\mu + i\omega_n) \hat{\mathbbm{1}} + \frac{i}{\tilde{r}^2}\left( 1 - i {\tilde{r}} {\rm sgn}(\omega_n) (\mu + i\omega_n) \right) {\tilde{\bf r}\cdot \hat{\sigma}} \right]I_{0}({\bf r}).
    \end{align}
    \item in mixed representation
    \begin{align}
        G^{\infty}(i\omega_n, {\bf k}_{\parallel}, z) = \frac{1}{2\pi}\int_{\mathbbm{R}} dk_z~\frac{i\omega_n + \hat{H}_{D}}{\sum_{i} v_i^2 k_i^2 - (\mu + i\omega_n)^2} e^{ik_z z} = \frac{1}{2\pi}\int_{\mathbbm{R}} dk_z~\frac{\mu + i\omega_n + v_x k_x \hat{\sigma}_x + v_y k_y \hat{\sigma}_y + v_z k_z \hat{\sigma}_z}{v_z^2 k_z^2 - ((\mu + i\omega_n)^2 - (v_x^2 k_x^2 + v_y^2 k_y^2))} e^{ik_z z},
    \end{align}
    which gives
    \begin{align}
        G^{\infty}(i\omega_n, {\bf k}_{\parallel}, z) = \frac{i}{2v_z} \left(  \frac{(\mu + i\omega_n) + v_x k_x \hat{\sigma}_x + v_y k_y \hat{\sigma}_y }{v_z q({\bf k}_{\parallel})} + {\rm sgn}(z) \hat{\sigma}_z\right) e^{i q({\bf k}_{\parallel}) |z|},
        \label{eq:gfdirac1D}
    \end{align}
    where $q({\bf k}_{\parallel}) = {\rm sgn}(\omega_n \mu) \sqrt{(\mu + i\omega_n)^2 - (v_x^2 k_x^2 + v_y^2 k_y^2)}\big/|v_z|$.
\end{enumerate}

\subsection{Junction Green's function}

In the case the location of cones in momentum space and the chemical potential change abruptly at $z = 0$, from $(v_L,\delta {\bf k}_L, \mu_L)$ for $z < 0$ to $(v_R,\delta {\bf k}_R, \mu_R)$ for $z > 0$, the Hamiltonian
\begin{align}
    \hat{H}_W &= \sum_{i = 1}^{2} v_i (k_i - \delta k_i(z)) \hat{\sigma}_i - \mu(z),
    \label{eq:surfacedirac}
\end{align}
can be studied separately for $z < 0$ and $z > 0$. For this reason, we introduce the Green's functions $G_{L}(i\omega_n, {\bf k}_{\parallel}, z, z^{\prime})$ and $G_{R}(i\omega_n, {\bf k}_{\parallel}, z, z^{\prime})$ defined for $z^{\prime} < 0$ and $z^{\prime} > 0$ respectively, \textit{i.e.} on the left and right side of the junction. 

The integration of the Weyl equation~\eqref{eq:surfacedirac} on a small segment around the interface, at $z = 0$, leads to the following boundary condition between $G^{L}$ and $G^{R}$
\begin{align}
    G^{L}(i\omega_n, {\bf k}_{\parallel}, z, 0) =  G^{R}(i\omega_n, {\bf k}_{\parallel}, z, 0).
    \label{eq:boundaryconditiondirac}
\end{align}
The Green's function of the junction can be expanded over the bulk Green's functions, $G_{L}^{\infty}$ and $G_{R}^{\infty}$, that are respectively the Green's functions for a bulk material with parameters $(v_L,\delta {\bf k}_L, \mu_L)$ and $(v_R,\delta {\bf k}_R, \mu_R)$. This is similar to method of images in electromagnetism and we use it to satisfy the boundary condition~\eqref{eq:boundaryconditiondirac}. In the following we illustrate this expansion for $z < 0$
\begin{align}
    &G_L(i\omega_n, {\bf k}_{\parallel}, z, z^{\prime}) = G^{\infty}_L(i\omega_n, {\bf k}_{\parallel}, z - z^{\prime}) + \hat{R}_{-} G^{\infty}_L(i\omega_n, {\bf k}_{\parallel}, z + z^{\prime}) = G^{\infty}_L(i\omega_n, {\bf k}_{\parallel}, z - z^{\prime}) + G^{\infty}_L(i\omega_n, {\bf k}_{\parallel}, z + z^{\prime}) \hat{R}_{+},\\
    &G_R(i\omega_n, {\bf k}_{\parallel}, z, z^{\prime}) = \hat{T}_{-} G^{\infty}_R(i\omega_n, {\bf k}_{\parallel}, z - z^{\prime}) = G^{\infty}_R(i\omega_n, {\bf k}_{\parallel}, z - z^{\prime}) \hat{T}_{+},
\end{align}
where we introduce the reflection and transmission operators $\hat{R}_{\pm}$ and $\hat{T}_{\pm}$. We introduce the notation $\pm$ in order to account for the fact that, in general, $\hat{R}$ and $\hat{T}$ do not commute with $G^{\infty}$.

We simplify our calculation by expanding the reflection and transmission operators as
\begin{align}
    R_{\pm} &= \alpha_{0\pm} \hat{\mathbbm{1}} + \alpha_{1\pm} {\bf h}_{\parallel,L}\cdot \hat{\sigma},\\
    T_{\pm} &= \big[ \beta_{0\pm} \hat{\mathbbm{1}} + \beta_{1\pm} {\bf h}_{\parallel,R}\cdot \hat{\sigma} \big]\frac{v_{z,L}}{v_{z,R}} e^{i(q_L({\bf k}_{\parallel}) - q_R({\bf k}_{\parallel}))z},
\end{align}
where ($\alpha_{0\pm},\alpha_{1\pm},\beta_{0\pm},\beta_{1\pm})$ are solve from the boundary condition~\eqref{eq:boundaryconditiondirac}. Here we also introduce ${\bf h}_{\parallel} = \sum_{i = 1}^{2} v_i( k_i - \delta k_i) {\bf e}_i$, the in-plane component of the Hamiltonian.

The boundary condition is satisfied for the following set of coefficients
\begin{align}
    \alpha_{0+} &= \frac{(\tilde{\bf h}_L - \tilde{\bf h}_R) \cdot \tilde{\bf h}_R - \tilde{q}_R i \tilde{\bf h}_L\times \tilde{\bf h}_R}{\tilde{\bf h}_L^2\tilde{\bf h}_R^2 - (1 + \tilde{q}_L \tilde{q}_R) \tilde{\bf h}_L\cdot\tilde{\bf h}_R + (\tilde{q}_L + \tilde{q}_R) i\tilde{\bf h}_L\times \tilde{\bf h}_R} \tilde{\bf h}_L^2 v_{z,L},\\
    \alpha_{1+} &= \frac{-i}{q_L(0)} \frac{(\tilde{\bf h}_L - \tilde{\bf h}_R) \cdot \tilde{\bf h}_R - \tilde{q}_R i \tilde{\bf h}_L\times \tilde{\bf h}_R}{\tilde{\bf h}_L^2\tilde{\bf h}_R^2 - (1 + \tilde{q}_L \tilde{q}_R) \tilde{\bf h}_L\cdot\tilde{\bf h}_R + (\tilde{q}_L + \tilde{q}_R) i\tilde{\bf h}_L\times \tilde{\bf h}_R} \tilde{\bf q}_Lv_{z,L},\\
    \beta_{0+} &= - \frac{(\tilde{q}_L + \tilde{q}_R )\tilde{\bf h}_L\cdot\tilde{\bf h}_R - (1 + \tilde{q}_L\tilde{q}_R) i \tilde{\bf h}_L\times \tilde{\bf h}_R}{\tilde{\bf h}_L^2\tilde{\bf h}_R^2 - (1 + \tilde{q}_L \tilde{q}_R) \tilde{\bf h}_L\cdot\tilde{\bf h}_R + (\tilde{q}_L + \tilde{q}_R) i\tilde{\bf h}_L\times \tilde{\bf h}_R} \tilde{\bf q}_Rv_{z,R},\\
    \beta_{1+} &= \frac{-i}{q_R(0)} \frac{(\tilde{\bf h}_L - \tilde{\bf h}_R) \cdot \tilde{\bf h}_L + \tilde{q}_L i \tilde{\bf h}_L\times \tilde{\bf h}_R}{\tilde{\bf h}_L^2\tilde{\bf h}_R^2 - (1 + \tilde{q}_L \tilde{q}_R) \tilde{\bf h}_L\cdot\tilde{\bf h}_R + (\tilde{q}_L + \tilde{q}_R) i\tilde{\bf h}_L\times \tilde{\bf h}_R} \tilde{\bf q}_Rv_{z,R},
\end{align}
or for
\begin{align}
    \alpha_{0-} &= \frac{(\tilde{\bf h}_L - \tilde{\bf h}_R) \cdot \tilde{\bf h}_R + \tilde{q}_R i \tilde{\bf h}_L\times \tilde{\bf h}_R}{\tilde{\bf h}_L^2\tilde{\bf h}_R^2 - (1 + \tilde{q}_L \tilde{q}_R) \tilde{\bf h}_L\cdot\tilde{\bf h}_R - (\tilde{q}_L + \tilde{q}_R) i\tilde{\bf h}_L\times \tilde{\bf h}_R} \tilde{\bf h}_L^2 v_{z,L},\\
    \alpha_{1-} &= \frac{i}{q_L(0)} \frac{(\tilde{\bf h}_L - \tilde{\bf h}_R) \cdot \tilde{\bf h}_R + \tilde{q}_R i \tilde{\bf h}_L\times \tilde{\bf h}_R}{\tilde{\bf h}_L^2\tilde{\bf h}_R^2 - (1 + \tilde{q}_L \tilde{q}_R) \tilde{\bf h}_L\cdot\tilde{\bf h}_R - (\tilde{q}_L + \tilde{q}_R) i\tilde{\bf h}_L\times \tilde{\bf h}_R} \tilde{\bf q}_L v_{z,L},\\
    \beta_{0-} &= - \frac{(\tilde{q}_L + \tilde{q}_R )\tilde{\bf h}_L\cdot\tilde{\bf h}_R + (1 + \tilde{q}_L\tilde{q}_R) i \tilde{\bf h}_L\times \tilde{\bf h}_R}{\tilde{\bf h}_L^2\tilde{\bf h}_R^2 - (1 + \tilde{q}_L \tilde{q}_R) \tilde{\bf h}_L\cdot\tilde{\bf h}_R - (\tilde{q}_L + \tilde{q}_R) i\tilde{\bf h}_L\times \tilde{\bf h}_R} \tilde{\bf q}_R v_{z,R}\\
    \beta_{1-} &= \frac{i}{q_R(0)} \frac{(\tilde{\bf h}_L - \tilde{\bf h}_R) \cdot \tilde{\bf h}_L - \tilde{q}_L i \tilde{\bf h}_L\times \tilde{\bf h}_R}{\tilde{\bf h}_L^2\tilde{\bf h}_R^2 - (1 + \tilde{q}_L \tilde{q}_R) \tilde{\bf h}_L\cdot\tilde{\bf h}_R - (\tilde{q}_L + \tilde{q}_R) i\tilde{\bf h}_L\times \tilde{\bf h}_R} \tilde{\bf q}_R v_{z,R}.
\end{align}
where the tilde notation implies that a quantity is normalized by $q_{0,L/R} = \mu_{L/R} + i\omega_n$ (for example $\tilde{\bf h}_L = {\bf h}_L/q_{0,L}$ or $\tilde{\bf h}_R = {\bf h}_R/q_{0,R}$). We also use the simplified notation ${\bf h}_L \times {\bf h}_R \equiv ({\bf h}_L \times {\bf h}_R)\cdot {\bf e}_z$. All these expressions are obtained for $z < 0$; if instead $z > 0$ then we should switch $i \rightarrow -i$.

The Fourier transforms of $G^{\mp}({\bf k}_{\parallel})$ to position-space are performed numerically and we use the expressions above to draw Figs.~2 and 3.

\subsection{Smooth junction Green's function in the WKB approximation}
\label{app:sec:wkb}

In the previous section we solved the Green's function of an abrupt interface between regions with different parameter $(v,\delta {\bf k}, \mu)$ in the Hamiltonian~\eqref{app:eq:hd}. We now consider the situation where the parameters smoothly vary from $(v,\delta {\bf k}, \mu)_L$ for $z \rightarrow -\infty$ to $(v,\delta {\bf k}, \mu)_R$ for $z\rightarrow \infty$. The parameters are supposed to be only $z$ dependent and in the following we solve for the Green's function of the Weyl equation~\eqref{app:eq:hd} in the WKB approximation following the method in~\cite{Pfirsch1991}.

We start by introducing the inverse of the thermal Green's function with smoothly varying parameters
\begin{align}
	\hat{H} &= h_x(z)\hat{\sigma}_x + h_y(z) \hat{\sigma}_y + \bigg( -i v(z) \partial_z - \frac{i}{2}v'(z) - \delta k_z(z) \bigg) \hat{\sigma}_z - (\mu(z) + i\omega_n + i {\rm sign}(\omega_n)/(2\tau(z)))\hat{\mathbbm{1}},
\end{align}
to which we search for an eigensolution of the form $\psi(z) = \Psi(z) e^{iS(z)}$ where the amplitude $\Psi(z)$ varies slowly compared to $e^{iS(z)}$. This then satisfies
\begin{align}
	\left( h_x(z)\hat{\sigma}_x + h_y(z) \hat{\sigma}_y + v(z) \bigg( k - \delta k_z(z)\bigg)\hat{\sigma}_z - (\tilde{\mu}(z) + \lambda )\hat{\mathbbm{1}} \right) \Psi(z) = 0,
	\label{app:eq:jacobi}
\end{align}
where $\tilde{\mu}(z) = \mu(z) + i\omega_n + i {\rm sign}(\omega_n)/(2\tau(z))$,  $k =\partial S/\partial z$. We can diagonalize this equation on the states $\phi_{\sigma = \pm}$ defined by the spinors
\begin{align}
	\phi_{\sigma} = \frac{1}{\sqrt{2}}
	\left(
	\begin{array}{c}
		\left( 1 + \sigma \frac{h_z}{\delta h} \right)^{1/2}\\
		\sigma\left( 1 - \sigma \frac{h_z}{\delta h} \right)^{1/2} e^{i\theta},
	\end{array}
	\right)
\end{align}
with $e^{i\theta} = (h_x+i h_y)/|h_{\parallel}|$ which leads to the projectors
\begin{align}
	\phi_{\sigma}^{\dagger}(z^{\prime})\phi_{\sigma}(z) = \frac12
	\left(
		\begin{array}{cc}
			\left( 1 + \sigma \frac{h_z}{\delta h} \right)^{1/2}\left( 1 + \sigma \frac{h_z^{\prime}}{\delta h^{\prime}} \right)^{1/2} & 
			\sigma\left( 1 + \sigma \frac{h_z}{\delta h} \right)^{1/2}\left( 1 - \sigma \frac{h_z^{\prime}}{\delta h^{\prime}} \right)^{1/2} e^{-i\theta^{\prime}} \\
			\sigma\left( 1 - \sigma \frac{h_z}{\delta h} \right)^{1/2}\left( 1 + \sigma \frac{h_z^{\prime}}{\delta h^{\prime}} \right)^{1/2} e^{i\theta}
			&
			\left( 1 - \sigma \frac{h_z}{\delta h} \right)^{1/2}\left( 1 - \sigma \frac{h_z^{\prime}}{\delta h^{\prime}} \right)^{1/2} e^{i(\theta - \theta^{\prime})}
		\end{array}
	\right),
\end{align}
where $\delta h = (h_{\parallel}^2 + h_z^2)^{1/2}$ with $h_z = v_z(\partial_z S - \delta k_z)$. We then have for each eigensolution of Eq.~\eqref{app:eq:jacobi}
\begin{align}
	&\frac{\partial S_{\sigma}}{\partial z} = k_{\sigma} = \delta k_z(z) + \frac{h(z)}{v_z(z)} = \delta k_z + \frac{1}{v_z}\left((\tilde{\mu} + \lambda_{\sigma}) - h_{\parallel}^2\right)^{1/2}\\
	\implies &S(z) = \int^{z}_{z_0} du~\left( \delta k_z(u) + \frac{1}{v_z(u)}((\tilde{\mu}(u) + \lambda_{\sigma})^2 - h_{\parallel}^2(u) )^{1/2}\right).
\end{align}
The phase factor $S(z)$ is defined up to a constant of integration, associated to the starting position $z_0$, and that we replace with parameter
\begin{align}
	a_{\sigma} \equiv k_{\sigma}(z_0) = \delta k_0 + \frac{1}{v_{z,0}} ( (\tilde{\mu}_0 + \lambda_{\sigma})^2 - h_{\parallel,0}^2)^{1/2},
\end{align}
and write the eigenvalues $\lambda_{\sigma}$ as a function of $a_{\sigma}$
\begin{align}
	\lambda_{\sigma} = - \tilde{\mu}_0 + \sigma \delta h_0 = -\tilde{\mu}_0 + \sigma ( h_{x,0}^2 + h_{y,0}^2 + v_{z,0}^2(a_{\sigma}^2 - \delta k_0)^2 )^{1/2}.
\end{align}
In this approximation we discard the trajectories that cycle, \textit{i.e.} the trajectory is uniquely defined by the parameter $a$. Also, since there is a one-to-one correspondence between $\lambda$ and $a$, we can use either parameter to define the trajectory.

It can be shown that the amplitude $\Psi(z)$ is given by the van Vleck determinant, usually related to the local density of states of semiclassical trajectories~\cite{Pfirsch1991}. When parametrizing the trajectory with the eigenvalue $\lambda$, we thus have
\begin{align}
	\Psi_{\sigma}(z) = \left(\frac{\partial S}{\partial z \partial \lambda_{\sigma}}\right)^{1/2} = \left( \frac{1}{v_z(z)} \frac{\tilde{\mu}(z) + \lambda_{\sigma}}{\left( (\tilde{\mu}(z) + \lambda_{\sigma})^2 - h_{\parallel}^2(z)\right)^{1/2}} \right)^{1/2}.
\end{align}

Then the WKB Green's function is obtained by integrating the previous solution over all eigenvalues $\lambda$
\begin{align}
	G(i\omega_n, {\bf k}_{\parallel}, z, z^{\prime}) = \frac{1}{2}\sum_{\sigma} \int \frac{d\lambda_{\sigma}}{2\pi} ~\frac{1}{\lambda_{\sigma}} \phi_{\sigma}^{\dagger}(z^{\prime})\phi_{\sigma}(z) \Psi_{\sigma}(z^{\prime}) \Psi_{\sigma}(z) e^{i\int^{z^{\prime}}_{z} du~\left( \delta k_z(u) + \frac{1}{v_z(u)}((\tilde{\mu}(u) + \lambda_{\sigma})^2 - h_{\parallel}^2(u) )^{1/2}\right)},
\end{align}
where the factor $1/2$ appears because we consider a bi-spinor~\cite{Pfirsch1991}. We can compute the above integral with the Cauchy theorem by choosing a contour such that the integrrand is negligible at large $|\lambda|$, that is such that ${\rm Im}(S(\lambda)) > 0 \implies \partial S/\partial z \sim {\rm sign}((z^{\prime}-z) v_z(u) \mu(u) \omega_n)$. We obtain
\begin{align}
    &G(i\omega_n, {\bf k}_{\parallel}, z, z^{\prime}) = \frac{i}{2} e^{i\int_{z}^{z^{\prime}} du~\left( \delta k_z(u) + {\rm sign}\left[(z^{\prime}-z)\omega_n v_z(u) \mu(u)\right] \frac{1}{v_z(u)} \left(\tilde{\mu}^2(u) - h_{\parallel}^2(u)\right)^{1/2}\right)} \times\\
    &\left( 
        \begin{array}{cc}
            \left( a(z,z^{\prime})/v_z^{\prime}\right)^{1/2} \left(  b(z,z^{\prime})/v_z\right)^{1/2} &
            \left( 1/(a(z,z^{\prime})v_z^{\prime} h_z^{\prime 2} ) \right)^{1/2}\left( b(z,z^{\prime})/{v_z^{}} \right)^{1/2}\left(h_{x}^{\prime} - ih_{y}^{\prime}\right)\\
            \left( a(z,z^{\prime})/v_z^{\prime} \right)^{1/2}\left( 1/(b(z,z^{\prime})v_z^{} h_z^{2}) \right)^{1/2}\left(h_{x}^{} + ih_{y}^{}\right)
            & 
            \left( 1/(a(z,z^{\prime})v_z^{\prime} h_z^{\prime 2} ) \right)^{1/2}\left( 1/(b(z,z^{\prime})v_z^{} h_z^{2}) \right)^{1/2}\left(h_x^{\prime} - i h_y^{\prime}\right)\left(h_x + i h_y\right)
        \end{array}
    \right),\nonumber
\end{align}
where
\begin{align}
	&h_z(z) = {\rm sign}(\omega_n \mu(z) v_z(z))  \left( \tilde{\mu}^2(z) - h_{\parallel}^2(z) \right)^{1/2},\\
	&\tilde{\mu}(z) = \mu(z) + i\omega_n + i {\rm sign}(\omega_n)/(2\tau(z)),\\
	&a(z,z^{\prime}) = \frac{\tilde{\mu}_{z^{\prime}}}{h_z^{\prime}} + {\rm sign}(z^{\prime}-z),\\
	&b(z,z^{\prime}) = \frac{\tilde{\mu}_{z}}{h_z} + {\rm sign}(z^{\prime}-z).
\end{align}

Note that we cannot further simplify these expressions since the square root has a branch cut on $(-\infty,0]$ so $\sqrt{a b} = \sqrt{a} \sqrt{b}$ is only defined if
\begin{align}
	\begin{array}{lc}
		{\rm (1): } & {\rm Im}(a) {\rm Im}(b) \leq 0,\\
		{\rm (2): }& {\rm Im}(a) {\rm Im}(b) > 0~{\rm and}~{\rm Im}(a){\rm Im}(ab) > 0.
	\end{array}
\end{align}

\section{Current response in the bulk}

In this section we compute the non-local conductivity in various situation for a planar electric field ${\bf E}({\bf x}) = {\bf E}_0\delta(z - z_0)$ and measuring the current in a plane $j_{\mu}({\bf x}) = j_{\mu}(z_1) \delta(z-z_1)$, such that $j_{\mu}(z_1) = \sigma_{\mu\mu}(z_1,z_0) E_{0\mu}$.

\subsection{Non-local conductivity and time-reversal symmetry}

The conductivity tensor can always be decomposed over a symmetric, $\sigma_1$, and an antisymmetric, $\sigma_2$, component with respect to time-reversal symmetry~\cite{BarangerStone89}
\begin{align}
    \sigma_{1,\mu\nu} &= - \sigma_0 \int_{\mathbbm{R}} d\epsilon~ \left(-\frac{df}{d\epsilon}\right) {\rm Tr}\left[ \hat{J}_{\mu} \left(G_{+}({\bf x}, {\bf x}^{\prime}) - G_{-}({\bf x}, {\bf x}^{\prime})\right) \hat{J}_{\nu} \left(G_{+}({\bf x}^{\prime}, {\bf x}) - G_{-}({\bf x}, {\bf x})\right)\right]\\
    \sigma_{2,\mu\nu} &= \sigma_0 \int_{\mathbbm{R}} d\epsilon~ \left(-\frac{df}{d\epsilon}\right) \left\{ {\rm Tr}\left[ \hat{J}_{\mu} G_{+}({\bf x}, {\bf x}^{\prime}) \hat{J}_{\nu} G_{-}({\bf x}^{\prime}, {\bf x})\right] - {\rm Tr}\left[ \hat{J}_{\mu} G_{-}({\bf x}, {\bf x}^{\prime}) \hat{J}_{\nu} G_{+}({\bf x}^{\prime}, {\bf x})\right] \right\}\nonumber\\
    &- \sigma_0 \int_{\mathbbm{R}} d\epsilon~ f(\epsilon) \left\{ {\rm Tr}
    \left[ 
        \hat{J}_{\mu} \partial_{\epsilon} G_{+}({\bf x}, {\bf x}^{\prime}) \hat{J}_{\nu} G_{+}({\bf x}^{\prime}, {\bf x})\right] 
        + {\rm Tr}\left[ \hat{J}_{\mu} G_{-}({\bf x}, {\bf x}^{\prime}) \hat{J}_{\nu} \partial_{\epsilon} G_{-}({\bf x}^{\prime}, {\bf x})
    \right]\right.~\nonumber\\
    &\left.
    ~~~~~~~~~~~~~~~~~~~~~~~~~~~~  ~~
        -
        {\rm Tr}\left[\hat{J}_{\mu} \partial_{\epsilon} G_{-}({\bf x}, {\bf x}^{\prime}) \hat{J}_{\nu} G_{-}({\bf x}^{\prime}, {\bf x})\right] 
        - {\rm Tr}\left[ \hat{J}_{\mu} G_{+}({\bf x}, {\bf x}^{\prime}) \hat{J}_{\nu} \partial_{\epsilon} G_{+}({\bf x}^{\prime}, {\bf x})
    \right] 
    \right\},
\end{align}
with $G_{\pm} = (\hat{H} \pm i0^{+})^{-1}$. In general $\sigma_2$ is not zero but one can check that, for $\mu = \nu$, there is the symmetry
\begin{align}
    \sigma_{2,\mu\mu}({\bf x},{\bf x}^{\prime}) =  -\sigma_{2,\mu\mu}({\bf x}^{\prime},{\bf x}),
\end{align}
so
\begin{align}
    \sigma_{1,\mu\mu}({\bf x},{\bf x}^{\prime}) =  \frac12\left( \sigma_{\mu\mu}({\bf x},{\bf x}^{\prime}) + \sigma_{\mu\mu}({\bf x}^{\prime},{\bf x}) \right).
\end{align}
That is, longitudinal components of the conductivity tensor can be obtained by averaging the longitudinal conductivities obtained when permuting the two probes. In the main text and below we focus on $\sigma_{1,\mu\mu}$ and drop the reference to the index $1$.

\subsection{Kubo formalism}

The expression of the non-local conductivity is obtained in the Kubo formalism in~\cite{PhysRevB.51.10085,BarangerStone89}. In the situation of a planar electric field ${\bf E}({\bf x}) = {\bf E}_0\delta(z - z_0)$ and measuring the current in a plane $j_{\mu}({\bf x}) = j_{\mu}(z_1) \delta(z-z_1)$, such that $j_{\mu}(z_1) = \sigma_{\mu\mu}(z_1,z_0) E_{0\mu}$, we obtain
\begin{align}
    \sigma_{\mu\nu}(z,z^{\prime}) = \int \dfrac{dS_z dS_{z^{\prime}}}{\pi \mathcal{A}}{\rm Tr}\left( \hat{j}_{\mu} {\rm Im}\hat{G}(z,z^{\prime})\hat{j}_{\nu} {\rm Im}\hat{G}(z,z^{\prime})\right).
    \label{app:eq:sigma}
\end{align}
where the current operator for the Weyl equation is $\hat{\bf j} = v(z) \hat{\boldsymbol{\sigma}}$ and where ${\rm Im}\hat{G}(z,z^{\prime}) \equiv \frac12\left( \hat{G}(i0^+,{\bf r},{\bf r}^{\prime}) - \hat{G}(i0^-,{\bf r}, {\bf r}^{\prime}) \right)$ with
\begin{align}
    \hat{G}(i0^+,{\bf r},{\bf r}^{\prime}) - \hat{G}(i0^-,{\bf r}, {\bf r}^{\prime}) = \frac{1}{4\pi v^2 |{\bf r} - {\bf r}^{\prime}|}&\left[\tilde{\mu} e^{i \tilde{\mu} |{\bf r}-{\bf r}^{\prime}|/|v|} \hat{\mathbbm{1}} - \tilde{\mu}^* e^{-i \tilde{\mu}^* |{\bf r}-{\bf r}^{\prime}|/|v|}\hat{\mathbbm{1}} + i \frac{v ({\bf r} - {\bf r}^{\prime})}{|{\bf r} -{\bf r}^{\prime}|^2}\cdot \hat{\sigma}\left( e^{i \tilde{\mu} |{\bf r}-{\bf r}^{\prime}|/|v|} - e^{-i \tilde{\mu} |{\bf r}-{\bf r}^{\prime}|/|v|} \right)\right.\nonumber \\
    &\left.+ {\rm sign}(v)\frac{{\bf r}-{\bf r}^{\prime}}{|{\bf r}-{\bf r}^{\prime}|}\cdot \hat{\sigma} \left( \tilde{\mu} e^{i \tilde{\mu} |{\bf r}-{\bf r}^{\prime}|/|v|} + \tilde{\mu}^* e^{-i \tilde{\mu} |{\bf r}-{\bf r}^{\prime}|/|v|} \right) \right],
\end{align}
where we use the notation $\tilde{\mu} = \mu + i/(2\tau)$.
Since the Green's function is translation invariant in the $(x,y)$ plane, we can easily get rid of one of the surface integrals in Eq.~\eqref{app:eq:sigma}. We perform the second surface integral in polar coordinates and obtain
\begin{align}
    \label{eq:sigmaxxdiracn}
    \sigma_{xx} &= -\frac{2\hbar v_x^2}{(4\pi v^2)^2\pi} \int \frac{dR}{R} d\theta ~ \left[\left(\tilde{\mu} e^{i \tilde{\mu} R/|v|} - \tilde{\mu}^* e^{-i \tilde{\mu}^* R/|v|}\right)^2 + \frac{-X^2 + Y^2 + Z^2}{R^2} \left( \left(\tilde{\mu} + \frac{i|v|}{R}\right) e^{i \tilde{\mu} R/|v|} + \left(\tilde{\mu}^* - \frac{i|v|}{R}\right) e^{-i \tilde{\mu}^* R/|v|}\right)^2\right].
\end{align}
where $(X,Y,Z) = {\bf r}-{\bf r}^{\prime}$ and $X = R \cos(\theta)$, $Y = R \cos(\theta)$. This expression shows that $\sigma_{xx} = \sigma_{yy}$ so we can write $\sigma_{xx} = \frac12 (\sigma_{xx} + \sigma_{yy})$ and get rid of the angular integral by removing the $X^2 - Y^2$ term. We then have
\begin{align}
    \sigma_{xx} &= -\frac{4\hbar v_x^2}{(4\pi v^2)^2} \int \frac{dR}{R}~ \left[\left(\tilde{\mu} e^{i \tilde{\mu} R/|v|} - \tilde{\mu}^* e^{-i \tilde{\mu}^* R/|v|}\right)^2 + \frac{Z^2}{R^2} \left( \left(\tilde{\mu} + \frac{i|v|}{R}\right) e^{i \tilde{\mu} R/|v|} + \left(\tilde{\mu}^* - \frac{i|v|}{R}\right) e^{-i \tilde{\mu}^* R/|v|}\right)^2\right]\\
    &= -\frac{2\hbar v_x^2}{(4\pi v^2)^2} \int_{1}^{\infty} \frac{du}{u}~ \left[\left(\tilde{\mu} e^{i \tilde{\mu} Z u/|v|} - \tilde{\mu}^* e^{-i \tilde{\mu}^* Z u/|v|}\right)^2 + \frac{1}{u^2} \left( \left(\tilde{\mu} + \frac{i|v|}{Z u}\right) e^{i \tilde{\mu} Z u/|v|} + \left(\tilde{\mu}^* - \frac{i|v|}{Z u}\right) e^{-i \tilde{\mu}^* Z u/|v|}\right)^2\right]\\
    &= -\frac{4\hbar v_x^2}{(4\pi v^2)^2} \int_{1}^{\infty} {du}~ \left( \frac{\tilde{\mu}^2 e^{i (2\tilde{\mu} Z/|v|) u} + \tilde{\mu}^{*2} e^{i (2\tilde{\mu}^* Z/|v|) u} - 2|\tilde{\mu}|^2e^{i(\tilde{\mu}-\tilde{\mu}^*)Z/|v| u}}{u}\right.\nonumber\\
    &\left.+ \frac{\tilde{\mu}^2 e^{i (2\tilde{\mu} Z/|v|) u} + \tilde{\mu}^{*2} e^{i (2\tilde{\mu}^* Z/|v|) u} + 2|\tilde{\mu}|^2e^{i(\tilde{\mu}-\tilde{\mu}^*)Z/|v| u}}{u^3}  + \frac{2i|v|}{Z}\frac{\tilde{\mu} e^{2i\tilde{\mu} Z/|v| u} - \tilde{\mu}^* e^{-2i\tilde{\mu}^* Z/|v| u} + (\tilde{\mu}^* - \tilde{\mu})e^{i(\tilde{\mu}-\tilde{\mu}^*) Z/|v| u}}{u^4}\right.\nonumber\\
    &\left. - \frac{v^2}{Z^2} \frac{e^{2i\tilde{\mu} Z/|v| u} + e^{-2i\tilde{\mu}^* Z/|v| u} - 2e^{i(\tilde{\mu}-\tilde{\mu}^*) Z/|v| u}}{u^5}\right)\\
    &= -\frac{4\hbar v_x^2}{(4\pi v^2)^2}\left[ \tilde{\mu}^2(E_1(-2i\tilde{\mu} Z/|v|) + E_3(-2i\tilde{\mu} Z/|v|)) + \tilde{\mu}^{*2}(E_1(2i\tilde{\mu}^* Z/|v|) + E_3(2i\tilde{\mu}^* Z/|v|))\right.\nonumber \\
    &\left.+ 2|\tilde{\mu}|^2(E_3(-i(\tilde{\mu}-\tilde{\mu}^*)Z/|v|) - E_1(-i(\tilde{\mu}-\tilde{\mu}^*)Z/|v|)) + \frac{2i|v|}{Z}\left(\tilde{\mu} E_4(-2i\tilde{\mu} Z/|v|) - \tilde{\mu}^* E_{4}(2i\tilde{\mu}^* Z/|v|) \right)\right.\nonumber\\
    &\left.+ (\tilde{\mu}^* - \tilde{\mu}) E_4(-i(\tilde{\mu}-\tilde{\mu}^*)Z/|v|)) - \frac{v^2}{Z^2}\left(E_5(-2i\tilde{\mu} Z/|v|) + E_5(2i\tilde{\mu}^* Z/|v|) - 2E_5(-i(\tilde{\mu} -\tilde{\mu}^*) Z/|v|)\right)\right].
\end{align}
We can also compute $\sigma_{zz}$ as a function of $Z = |z - z^{\prime}|$ and obtain
\begin{align}
    &\sigma_{zz} = -\frac{2\hbar v_x^2}{(4\pi v^2)^2} \int \frac{dR}{R}~ \left[\left(\tilde{\mu} e^{i \tilde{\mu} R/|v|} - \tilde{\mu}^* e^{-i \tilde{\mu}^* R/|v|}\right)^2 + \left(1 - 2\frac{Z^2}{R^2}\right) \left( \left(\tilde{\mu} + \frac{i|v|}{R}\right) e^{i \tilde{\mu} R/|v|} + \left(\tilde{\mu}^* - \frac{i|v|}{R}\right) e^{-i \tilde{\mu}^* R/|v|}\right)^2\right]\\
    &= -\frac{2\hbar v_x^2}{(4\pi v^2)^2} \int_{1}^{\infty} \frac{du}{u}~ \left[\left(\tilde{\mu} e^{i \tilde{\mu} Z/|v| u} - \tilde{\mu}^* e^{-i \tilde{\mu}^* Z/|v| u}\right)^2 + \left(1 - \frac{2}{u^2}\right) \left( \left(\tilde{\mu} + \frac{i|v|}{Z u}\right) e^{i \tilde{\mu} Z/|v| u} + \left(\tilde{\mu}^* - \frac{i|v|}{Z u}\right) e^{-i \tilde{\mu}^* Z/|v| u}\right)^2\right]\\
    &= -\frac{2\hbar v_x^2}{(4\pi v^2)^2} \int_{1}^{\infty} du~ \left( \left({\tilde{\mu}^2e^{2i\tilde{\mu} Z/|v| u} + \tilde{\mu}^{*2}e^{-2i\tilde{\mu}^* Z/|v| u} }\right)\left(\frac{2}{u} - \frac{2}{u^3}\right) - \frac{4\tilde{\mu}\tilde{\mu}^* e^{-i(\tilde{\mu}- \tilde{\mu}^*) Z/|v| u}}{u^3}\right.\nonumber\\
    & + \frac{2i|v|}{Z}\left({\tilde{\mu} e^{i\tilde{\mu} Z/|v| u} - \tilde{\mu}^* e^{-i\tilde{\mu}^* Z/|v| u} + (\tilde{\mu}^* - \tilde{\mu}) e^{i(\tilde{\mu} - \tilde{\mu}^*)Z/|v| u}}\right)\left(\frac{1}{u^2} - \frac{2}{u^4}\right)\nonumber\\
    &\left.- \frac{v^2}{Z^2}\left({e^{2i\tilde{\mu} Z/|v| u} + e^{-2i\tilde{\mu}^* Z/|v| u} - 2e^{i\delta\tilde{\mu} Z/|v| u}}\right)\left(\frac{1}{u^3} - \frac{2}{u^5}\right)  \right)\\
    &= -\frac{2\hbar v_x^2}{(4\pi v^2)^2} \left( 2\tilde{\mu}^2( E_1(-2i\tilde{\mu} Z/|v|) - E_3(-2i\tilde{\mu} Z/|v|)) + 2\tilde{\mu}^{*2}(E_1(2i\tilde{\mu}^* Z/|v|) - E_3(2i\tilde{\mu}^* Z/|v|)) - 2|\tilde{\mu}|^2 E_3(-i(\tilde{\mu}-\tilde{\mu}^*)Z/|v|) \right.\nonumber\\
    &\left. +\frac{2i|v|}{Z}\left[\tilde{\mu}(E_2(2i\tilde{\mu} Z/|v|) - 2E_4(2i\tilde{\mu} Z/|v|)) - \tilde{\mu}^*(E_2(-2i\tilde{\mu}^* Z/|v|) - 2E_4(-2i\tilde{\mu}^* Z/|v|)) \right.\right.\nonumber\\
    &\left.\left. + (\tilde{\mu}^*-\tilde{\mu})(E_2(-i(\tilde{\mu} -\tilde{\mu}^*)Z/|v|) - 2E_4(-i(\tilde{\mu} -\tilde{\mu}^*) Z/|v|))\right] - \frac{v^2}{Z^2}(E_3(-2i\tilde{\mu} Z/|v|) - 2E_5(-2i\tilde{\mu} Z/|v|)\right.\nonumber\\
    &\left. + E_3(2i\tilde{\mu}^* Z/|v|) - 2E_5(2i\tilde{\mu}^* Z/|v|) - 2E_3(-i(\tilde{\mu} -\tilde{\mu}^*) Z/|v|)) + 2 \times2E_5(-i(\tilde{\mu} -\tilde{\mu}^*) Z/|v|))\right).
\end{align}

We show the behaviour of the non-local conductivity obtained from these two expressions in Fig.~\ref{app:fig:3}. These expressions also match with our numerical results for the non-local conductivity far from the junction.

\subsection{Semiclassical limit.}
In the semiclassical limit the wave-like nature of quasiparticles is neglected such that one can approximate the exponentials with complex arguments with a functions with the same decay length and same volume integral but without the quantum oscillations
\begin{align}
    e^{i \alpha Z u} \rightarrow \frac{i {\rm Im}(\alpha)}{\alpha} e^{- {\rm Im}(\alpha) Z u}
\end{align}
from which we also deduce the semiclassical limits of terms involving powers of $1/Z = 1/|z-z^{\prime}|$ by integrating this equation over $u$ to have
\begin{align}
    \frac{e^{i \alpha Z u}}{Z^n} \rightarrow \left(\frac{\alpha}{i {\rm Im}(\alpha)}\right)^{n-1}\frac{e^{- {\rm Im}(\alpha) Z u}}{Z^n}.
\end{align}

This semi-classical limit is evaluated for a non-relativistic electron gas in Ref.~\cite{PhysRevB.51.10085}, and for a Weyl electron gas we obtain
\begin{align}
    \sigma_{xx}^{\rm cl.}(Z) &= \frac{\hbar v_x^2}{(4\pi v^2)^2} \left(  \left({\rm Re}(\tilde{\mu})^2 + 2 {\rm Im}(\tilde{\mu})^2\right) E_1(2 {\rm Im}(\tilde{\mu}) Z/|v|) - {\rm Re}(\tilde{\mu})^2 E_3(2 {\rm Im}(\tilde{\mu}) Z/|v|)\right),\\
    \sigma_{zz}^{\rm cl.}(Z) &= \frac{2\hbar v_x^2}{(4\pi v^2)^2} \left(  {\rm Im}(\tilde{\mu})^2 E_1(2 {\rm Im}(\tilde{\mu}) Z/|v|) + {\rm Re}(\tilde{\mu})^2 E_3(2 {\rm Im}(\tilde{\mu}) Z/|v|)\right),
\end{align}
that indeed match the conductivity in the Kubo formula but without the oscillating behaviour (see Fig.~\ref{app:fig:3}). 

\begin{figure*}[htb]
	\centering
	\includegraphics[width=\textwidth]{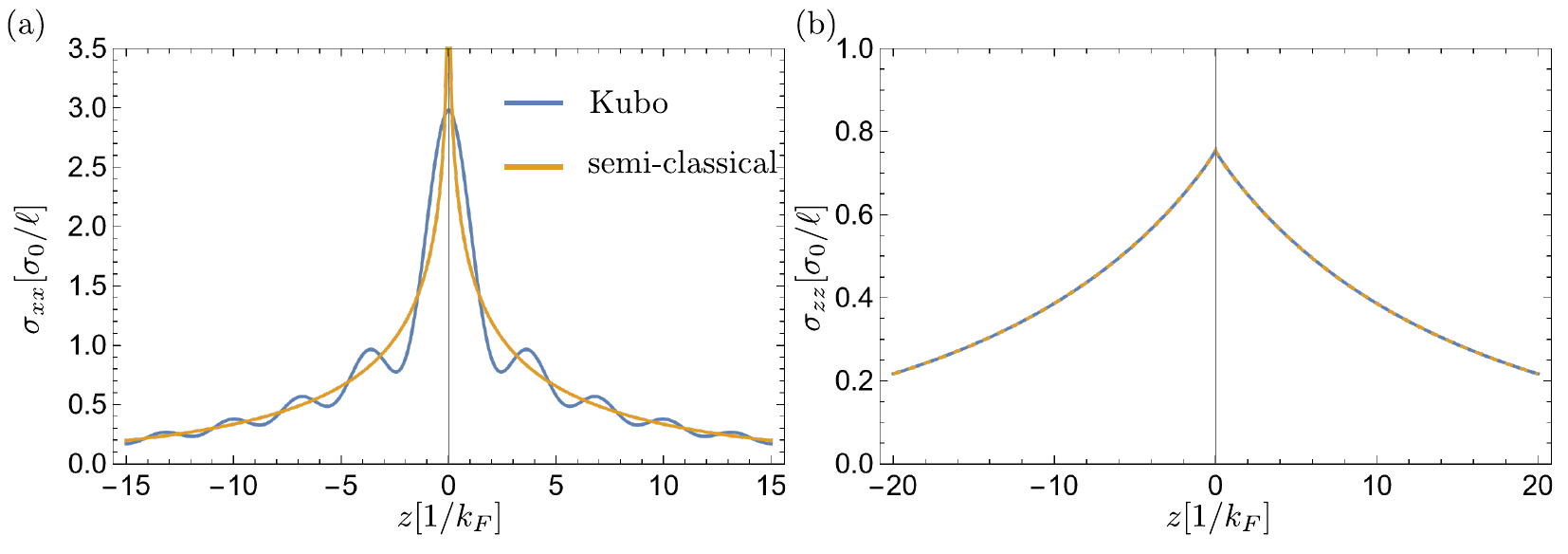}
	\caption{Bulk non-local conductivity of the Weyl electron gas for a planar electric field at $z = 0$. (a) $\sigma_{xx}$ and (b) $\sigma_{zz}$ are obtained for $(v_F, \mu, \ell) = (1, 1, 50)$, and are shown in units of the total bulk conductivity $\sigma_0$. We show the non-local conductivities from the Kubo formula in blue, and from the semi-classical approximation in orange, the two expression match relatively well up to the oscillations in $\sigma_{xx}$.}
	\label{app:fig:3}
\end{figure*}

\subsection{Total conductivity}

The global conductivity tensor is obtained after performing the integral over all the volume
\begin{align}
    \langle \sigma_{\mu\mu} \rangle = \frac{1}{L_z} \int dz dz^{\prime}~ \sigma_{\mu\mu}(z,z^{\prime}).
\end{align}
In the configuration with an electric field in the $z$ direction, we obtain
\begin{enumerate}
    \item For the non-relativistic electron gas:
    \begin{align}
        &\sigma_{xx} = \sigma_{yy} = \frac{1}{8\pi^2} \frac{{\rm Re}(q)^2 - m\mu/3}{{\rm Im}(q)} \frac{e^2}{\hbar},\\
        &\sigma_{zz} = \frac{1}{6\pi^2} \frac{\mu}{{\rm Im}(q)} \frac{e^2}{\hbar}
    \end{align}
    so the total conductivity is 
    \begin{align}
        \sigma_0 = \frac13 ( \sigma_{xx} + \sigma_{yy} + \sigma_{zz} ) = \frac{1}{12\pi^2} \frac{{\rm Re}(q)^2}{{\rm Im}(q)} \frac{e^2}{\hbar}
    \end{align}
    which is obtained in Ref.~\cite{PhysRevB.51.10085}.
    \item For the Weyl electron gas:
    \begin{align}
        &\sigma_{xx} = \sigma_{yy} = \sigma_{zz} = \frac{3 + 4 \mu^2\tau^2}{6\pi^2 |v| \tau} \frac{e^2}{\hbar}
    \end{align}
    so the total conductivity is 
    \begin{align}
        \sigma_0 = \frac13 ( \sigma_{xx} + \sigma_{yy} + \sigma_{zz} ) = \frac{3 + 4 \mu^2\tau^2}{6\pi^2 |v| \tau} \frac{e^2}{\hbar}.
    \end{align}
\end{enumerate}

\section{Conditions for optimal Veselago lensing}

In this section we explore how the smoothness of a $pn$ junction and how the mismatch in carrier densities affect the amplitude of Veselago lensing. We also show devices that realize a chiral Veselago lens and how they can be used to explore Veselago lensing from local sources.

\subsection{Veselago lensing for a smooth junction}

In the main text we discuss the situation of a sharp $pn$ junction, based on the absence of a built-in potential in chiral Veselago lens. In the situation of a smooth junction, with a characteristic length $d$ we instead have the behaviour reproduced in Fig.~\ref{app:fig:0}, obtained for $k_F d = 3/2$ with other parameters as in the main text $v_{F,L} = v_{F,R} = 1$, $\mu_{L} = -\mu_{R} = 1$ and for a mean free path $\ell = 50/k_F$. We see that the amplitude of the Veselago image is strongly reduced compared to the image in an sharp interface, even for such a small value of $d$.
\begin{figure*}[htb]
	\centering
	\includegraphics[width=\textwidth]{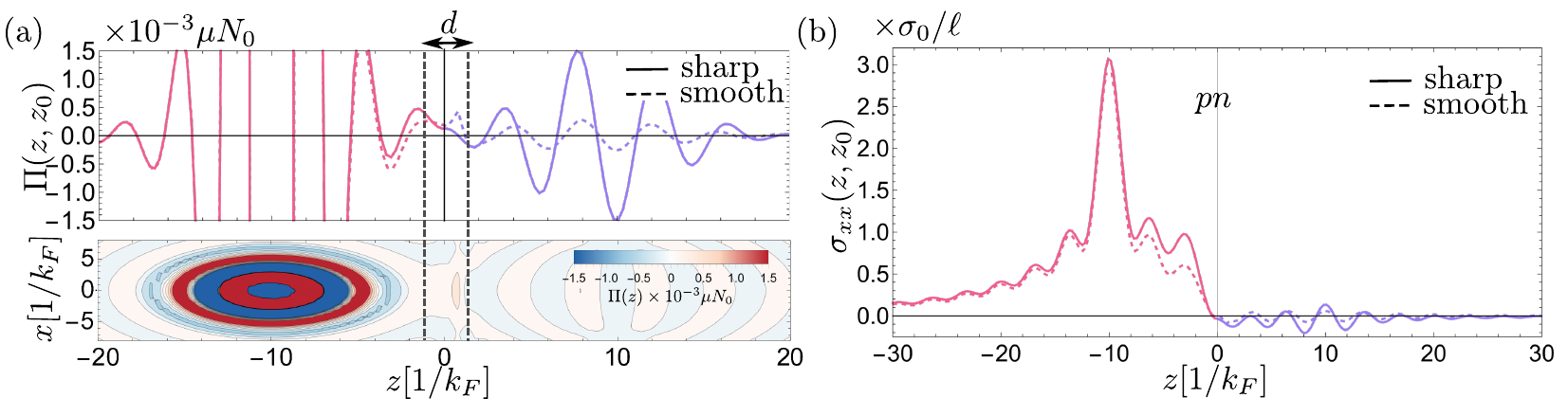}
	\caption{(a) Friedel oscillations $\Pi(z,z^{\prime})$ and (b) non-local conductivity $\sigma_{xx}$ of a $pn$ junction for a smooth $pn$ junction, with extension $d = 3/(2k_F)$. The density plot in (a) corresponds to the polarisability of the smooth $pn$ junction of the Weyl electron gas, obtained from the WKB approximation (see Sec.~\ref{app:sec:wkb}).}
	\label{app:fig:0}
\end{figure*}

\subsection{Veselago lensing for different densities}

In Figs.~\ref{app:fig:1}(a-c) we show the non-local conductivity $\sigma(z,z^{\prime})$ as a function of the location of the input potential different, at $z$, and the location of the output current, at $z^{\prime}$, for different values of the Fermi wavevector for $z > 0$ (see Fig.~2(d) in the main text).

We find that image current through the Veselago lens is largest when the carrier density on each side of the interface is the same (see Fig.~\ref{app:fig:1}(a)) and this is the situation we choose to illustrate in the main text. A $pn$ junction with different carrier densities still shows a peaked conductivity, due to Veselago lensing, but the amplitude of this peak decays faster than for identical carrier densities.

\begin{figure*}[htb]
	\centering
	\includegraphics[width=\textwidth]{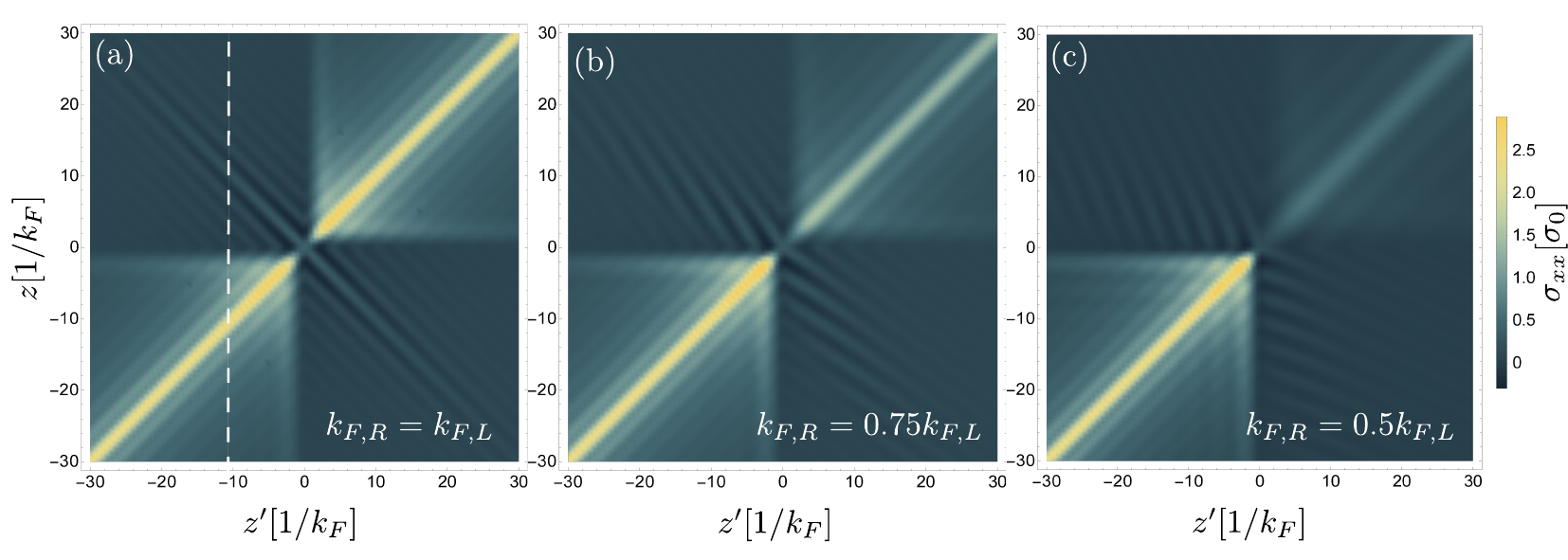}
	\caption{Non-local conductivity $\sigma_{xx}$ of a $pn$ junction with different carrier densities in the regions $z < 0$ and $z > 0$, given by the Fermi wavevectors $k_{F,L}$ and $k_{F,R}$ for (a) $k_{F,R} = k_{F,L}$, (b) $k_{F,R} = 0.75 k_{F,L}$ and (c) $k_{F,R} = 0.5 k_{F,L}$. The non-local conductivity along the dashed line in (a) corresponds to the data in Fig.~2(f) of the main text. }
	\label{app:fig:1}
\end{figure*}

\subsection{Devices to probe chiral Veselago lensing}

In the main text we consider that the chiral anomaly is induced locally from an extended electrode that applies a uniform electric field for $z < 0$, in presence of a uniform magnetic field. This is depicted in Fig.~1 of the main text, where we consider that the extended electric field acts as a source for current.

Other configurations are also possible to explore Veselago lensing as a consequence of the chiral anomaly and that we illustrate in Fig.~\ref{app:fig:2}(a,b). In both figures the principle is the same, we consider that the input voltage or electric field is applied on the surface normal to the extended electric field responsible for the chiral anomaly. 

\begin{figure*}[htb]
	\centering
	\includegraphics[width=\textwidth]{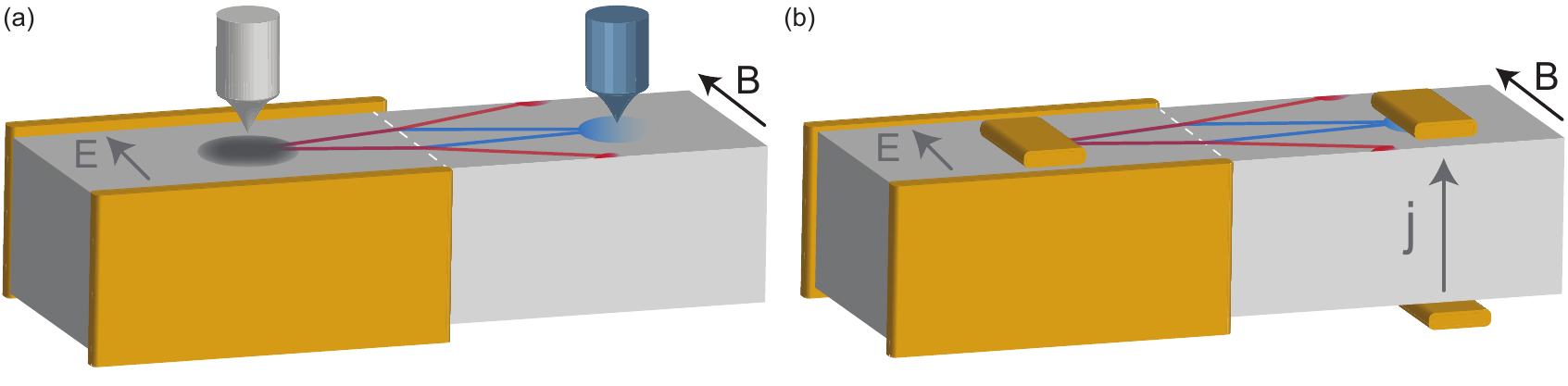}
	\caption{Devices to probe chiral Veselago lensing. (a) A local potential is applied with an STM tip (or impurity) on the surface of the device and leads to charge fluctuations that are measured by a second tip. (b) A planar electric field is applied on the surface of the device and leads to a current distribution that resembles that in Fig.~2(e,f) of the main text. In both (a) and (b), the side electrodes apply an electric field at $z<0$ that is colinear with the magnetic field, inducing a $pn$-junction, and thus a Veselago lens, for a single chirality.}
	\label{app:fig:2}
\end{figure*}

\end{widetext}

\end{document}